\documentclass[twocolumn,amsmath,amssymb,aps]{revtex4-2}
\usepackage{lineno,hyperref}
\usepackage{bm}
\usepackage{amsmath}
\usepackage{gensymb}
\usepackage{epstopdf}
\usepackage{booktabs}

\usepackage{graphicx}
\usepackage{ulem}
\usepackage{xcolor}
\usepackage{hyperref}
\usepackage{multirow}
\usepackage{csquotes}
\begin{document}

\title{Experimental observation of spin glass state in highly disordered quaternary Heusler alloy FeRuMnGa}

\author{Shuvankar Gupta$^{1,4}$}
\email{guptashuvankar5@gmail.com}
\author{Sudip Chakraborty$^{1,4}$}
\author{Santanu Pakhira$^2$}
\email{spakhira@ameslab.gov}
\author{Anis Biswas$^2$}
\author{Yaroslav Mudryk$^2$}
\author{Amit Kumar$^{3,4}$}
\author{Bodhoday Mukherjee$^5$}
\author{Gunadhor S. Okram$^5$}
\author{Amitabh Das$^{3,4}$}
\author{Vitalij K. Pecharsky$^{2,6}$}
\thanks{deceased}
\author{Chandan Mazumdar$^1$}

\affiliation{$^1$Condensed Matter Physics Division, Saha Institute of Nuclear Physics, 1/AF, Bidhannagar, Kolkata 700064, India}
\affiliation{$^2$Ames National Laboratory, Iowa State University, Ames, Iowa 50011, USA}
\affiliation{$^3$Solid State Physics Division, Bhabha Atomic Research Centre, Mumbai 400 085, India}
\affiliation{$^4$Homi Bhabha National Institute, Training School Complex, Anushaktinagar, Mumbai 400094, India}
\affiliation{$^5$UGC-DAE Consortium for Scientific Research, Universisty Campus, Khandwa Road Indore-452001, India}
\affiliation{$^6$Department of Materials Science and Engineering, Iowa State University, Ames, Iowa 50011, USA}

\date{\today}% It is always \today, today,
\begin{abstract}
The realization of spin-glass (S-G) state in Heusler alloys is very rare despite the presence of inherent structural and elemental disorder in those compounds. Although a few half and full Heusler alloys are known to exhibit S-G state, there is hardly any manifestation of the same in cases of quaternary Heusler compounds. Here we report the observation of S-G state in a highly disordered equiatomic quaternary Heusler compound: FeRuMnGa, where  the  S-G state is in between of canonical S-G and cluster glass. Different intricate features of S-G state including  non-equilibrium magnetic dynamics at low temperature in the compound are  unveiled through our comprehensive magnetic, heat capacity and neutron diffraction studies. The structural disorder in the sample is neither conventional \textit{A2}- nor \textit{B2}-type while those two types are commonly observed for Heusler compounds.  The presence of disorder also plays a significant role in electron transport properties of the alloy, which is reflected in its exhibition of semi-metallic behavior and anomalous Hall effect at low temperature.

\end{abstract}
\maketitle

\section{\label{sec:Introduction}Introduction}

In the field of material science and condensed matter physics, Heusler alloys continue to hold the pole position, even after 100 years of their discovery. With the passage of time, those materials remain in focus of intense study in various fields of research starting from thermoelectric~\cite{hinterleitner2019thermoelectric,mondal2018ferromagnetically}, magneto-caloric~\cite{liu2012giant,liu2019reversible}, spintronics~\cite{felser2007spintronics,bainsla2016equiatomic}, topological insulators~\cite{manna2018heusler,chadov2010tunable}, \textit{etc.} to the recently discovered magnetic skyrmions~\cite{saha2019intrinsic,madduri2020ac}. Generally, Heusler alloys are of two types: i) full Heusler represented as X$_2$YZ and ii) half Heusler represented as XYZ, where X and Y are the transition elements and Z is the \textit{sp}-group element~\cite{graf2011simple}. Recently, another new variant of Heulser alloy, named quaternary Heusler (X$X^{\prime}$YZ) alloy, was introduced~\cite{alijani2011electronic}. Most of the reported half-Heusler alloys contain only a single magnetic ion (Y) (mainly the Mn atom or rare-earth compounds), occupying the octahedral position~\cite{felser2015basics}. In contrast, the full Heusler compounds can have two different magnetic atoms (X, Y) occupying tetrahedral and octahedral  lattice positions, respectively~\cite{graf2013magnetic}. In such systems, besides the more localised Y-atoms (mainly the Mn atoms with more localized electrons), an additional delocalised sublattice containing of X-atoms also starts to develop. Heusler compounds of X$_2$YZ type can exhibit a wider variety of magnetic properties; \textit{viz}, ferromagnetism, ferrimagnetism, antiferromagnetism, half-metallic ferromagnetism (HMF) \textit{etc.}~\cite{graf2013magnetic}. Half-metallic ferromagnets (HMFs) are the special kind of material in which one sub-band behaves like a metal while the other sub-band behaves like semiconductor~\cite{de1983new}. Since the discovery of HMF nature in NiMnSb~\cite{de1983new}, Heusler alloys, in general, have drawn considerable interest of the spintronics community~\cite{felser2007spintronics}. In the vast family of Heusler alloys, the spintronic related research primarily focuses on the Co-based alloys that are known to exhibit a strong spin polarization and a relatively high Curie temperature~\cite{bombor2013half,graf2011simple}, while a scant attention is paid to other systems. The total magnetic moment for a ferromagnetic/ferrimagnetic full and quaternary Heusler alloy may be calculated using the Slater-Pauling (S-P) formula as m = (N$_V$ $-$ 24) $\mu_B$/f.u, where N$_V$ is the total number of valence electrons in the primitive cell. All Heulser-based  HMFs are known to follow S-P rule~\cite{galanakis2002slater,ozdougan2013slater,graf2011simple}. However, when a system forms with structural disorder, the magnetic interaction strength is impeded, although the compound often remains ferromagnetic. This weakened magnetic interaction strength usually leads to lower Curie temperatures as well as reduced value of saturation magnetic moment, violating the S-P rule~\cite{kanomata2010magnetic,fujita1972magnetic,wurmehl2006co2crin,mukadam2016quantification,rani2022disorder,rani2019experimental}. It is however not yet clear whether the structural disorder can indeed get rid of magnetic order completely, as there exist only a very few such studies concerning the Heusler alloy family~\cite{hiroi2009magnetic,hiroi2012spin,kroder2019spin}. The random variation of magnetic interaction strength caused by a strong structural disorder is expected to inhibit magnetic ordering in the system and may even introduce a reentrant spin-glass or even pure canonical/cluster glass state~\cite{chatterjee2009reentrant,ma2011coexistence,hiroi2009magnetic,hiroi2012spin,kroder2019spin}. Although there exist quite a few reentrant spin glass Heusler alloy systems, where the spin-glass state develops below their respective Curie temperatures~\cite{chatterjee2009reentrant,ma2011coexistence}, examples of a pure spin/cluster glass system are quite rare~\cite{hiroi2009magnetic,hiroi2012spin,kroder2019spin}. As of now, there is hardly any known quaternary Heusler alloy exhibiting clear a spin/cluster glass behaviour. In the present work, we report the structural and physical properties of FeRuMnGa, a new quaternary Heusler compound. Through different experimental techniques \textit{viz}. neutron diffraction, dc- and ac-susceptibility and different dynamical magnetic measurements, we demonstrate that the system forms with large atomic disorder and exhibits cluster spin-glass behaviour. The sample's structural disorder does not conform to either the conventional \textit{A2}- or \textit{B2}- types, which are commonly observed in Heusler compounds. In addition to this, the system shows non-metallic electron transport behavior.

\section{\label{sec:Methods}Experimental Methods}

Polycrystalline FeRuMnGa sample was prepared using arc melting technique in inert (argon) atmosphere taking appropriate high-purity ($>$99.9\%) constituent elements. The sample was melted 5 times, flipping after each melting for attaining better homogeneity. To compensate for the amount of Mn evaporated, an additional 2\% extra Mn was added during the melting. Room temperature powder X-ray diffraction (XRD) was performed using Cu-K$\alpha$ radiation in a TTRAX-III diffractometer (Rigaku Corp., Japan). The sample's single-phase nature was confirmed and crystal structure was determined from the XRD data by performing Rietveld refinement using the FullProf software package~\cite{rodriguez1993recent}. Magnetic properties were investigated using a SQUID magnetometer (Quantum Design Inc., USA) at temperatures ranging from 2 to 380 K and magnetic fields ranging from 0 to 70 kOe. For magnetic susceptibility measurements, both zero field-cooled (ZFC) and field-cooled (FC) methods were used. During the ZFC protocol, the sample was cooled to 2 K without the application of any external magnetic field, and magnetization measurements were performed in a specified magnetic field while heating from 2 to 380 K. In the FC procedure, the sample was cooled to 2 K in a magnetic field, and magnetization (M) \textit{versus} temperature (T) measurements were taken in the same field during heating. The isothermal magnetic-field dependence of magnetization, M \textit{versus} H, were measured at various temperatures. Before each series of M-H measurements, the sample was cooled from paramagnetic region to the required temperature in the absence of a magnetic field. AC susceptibility experiments were carried out in a 6 Oe excitation field with frequencies ranging from 1 to 1489 Hz. Heat capacity measurements were performed in standard relaxation method using Physical Property Measurement System (PPMS) (Quantum design Inc., USA). Neutron diffraction (ND) patterns on powdered sample were measured in the PD2 powder neutron diffractometer ($\lambda$= 1.2443 {\AA}) at the Dhruva reactor, Bhabha Atomic Research Centre (BARC), India. Electrical resistivity and magneto-transport measurements were  also carried out by conventional four probe method in the PPMS. A rectangular shaped sample was cut and polished for this purpose and silver epoxy was used for making electrical connections. Thermopower measurements were performed in the temperature range of 15$-$310 K using a home-built setup.

\section{Results and Discussion}

\subsection{\label{sec:XRD}X-ray diffraction}

\begin{figure}[h]
\centerline{\includegraphics[width=.48\textwidth]{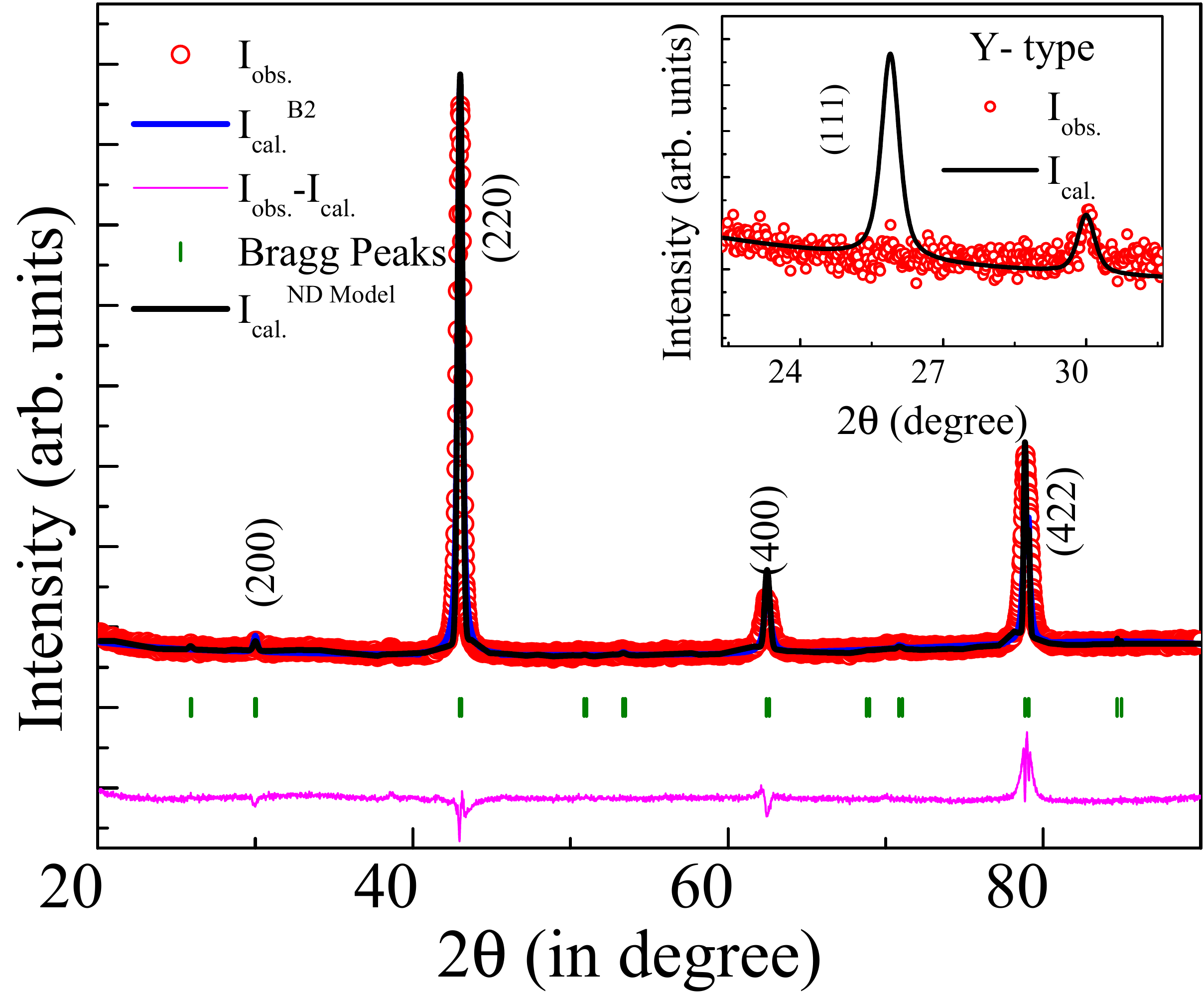}}
{\caption{Rietveld refinement of the powder XRD pattern of FeRuMnGa at room temperature. Blue line represent the fit using B2 type disorder, whereas the black line represent the fit using the same structural model employed to analyse the neutron diffraction data (Table~\ref{Occupancy_NPD_Table}). Miller indices for the corresponding  Bragg peaks are posted in brackets. Inset shows Rietveld refinement assuming ordered Y-type structure. Mismatch of the intensity at (111) peak is clearly evident.}\label{Fig_1}}
\end{figure}

Fig.~\ref{Fig_1} represents the XRD data of the as-prepared sample taken at room temperature. Our attempt to perform Rietveld refinement fit of the XRD data considering an ordered crystal structure (\textit{Y}-type, space group: \textit{F$\bar{4}$3m}, no. 216), ~\cite{bainsla2016equiatomic} in which Ga occupy 4\textit{a} (0,0,0), Mn 4\textit{b} (0.5,0.5,0.5), Fe 4\textit{c} (0.25,0.25,0.25) and Ru 4\textit{d} (0.75,0.75,0.75) atomic positions reveals a significant mismatch in the (111) peak intensity (inset of Fig.\ref{Fig_1}). It is worth mentioning here that the presence of (111) and (200) super-lattice peaks in the diffraction pattern is generally considered as an indication of ordered crystal structure in Heusler systems~\cite{gupta2022coexisting,venkateswara2015electronic}. However, as many Heusler alloys contain multiple elements from the same period of the periodic table with similar atomic sizes, the crystal structure often forms with atomic disorder~\cite{graf2011simple,balke2007structural}. The selective presence or absence of these two super-lattice peaks is indicative of the nature of such atomic disorder. For a quaternary Heusler alloy (X$X^{\prime}$YZ): assuming Z at 4\textit{a}, Y at 4\textit{b}, X at 4\textit{c} and $X^{\prime}$ at 4\textit{d}, the scattering factor for any random (\textit{hkl}) plane can be written as~\cite{rani2019spin}
\begin{equation}
F_{hkl} = 4(f_Z + f_{Y}e^{\pi{i(h+k+l)}} + f_{X}e^{{\frac {\pi}{2i}}(h+k+l)}+ f_{X^{\prime}}e^{{-\frac {\pi}{2i}}(h+k+l)})
\label{eq1}
\end{equation}
Accordingly, one can write the scattering factor for (111), (200) and (220) as
\begin{eqnarray}
F_{111} = 4(f_{Z} - f_{Y}) - i(f_X - f_{X^{\prime}})) \nonumber \\
F_{200} = 4[(f_Z + f_Y ) - (f_X + f_{X^{\prime}})]  \nonumber \\
F_{220} = 4[(f_Z + f_Y ) + (f_X + f_{X^{\prime}})]
\label{eq2}
\end{eqnarray}
\noindent
The two most frequently observed disorders in Heusler alloy are known as A2- and B2-types. In A2-type of disorder, all the elements (X, $X^{\prime}$, Y, Z) completely mix with each other in equivalent ratio and due to this random mixing, both the (111) and (200) peaks vanish from the diffraction pattern~\cite{graf2011simple,gupta2022coexisting}. For B2-type disorder, Y \& Z and X and $X^{\prime}$ atoms randomly mixes with each other in the 4\textit{a} \& 4\textit{b} and 4\textit{c} \&  4\textit{d} sites, respectively, giving rise to only (200) peak in the diffraction data. In the studied compound, the (111) is absent and the (200) peak is present in the XRD data suggesting presence of B2 type of disorder. The Rietveld refinement of the XRD data assuming B2-type of structure is presented in Fig.\ref{Fig_1}. The random mixing between Ga \& Mn and Fe \& Ru in the 4\textit{a} \& 4\textit{b} and 4\textit{c} \& 4\textit{d} sites, respectively, fits the experimental data quite satisfactorily. The lattice parameter is estimated to be = 5.935 \AA.  A further refinement of structural disorder has been carried out using neutron diffraction experiment and presented later in Sec.~\ref{sec:Neutron}. The corresponding Rietveld refinemnet assuming this structural disorder has been also presented in Fig.~\ref{Fig_1}.

\subsection{\label{sec:Magnetism}dc magnetization study}

\begin{figure}[h]
\centerline{\includegraphics[width=.48\textwidth]{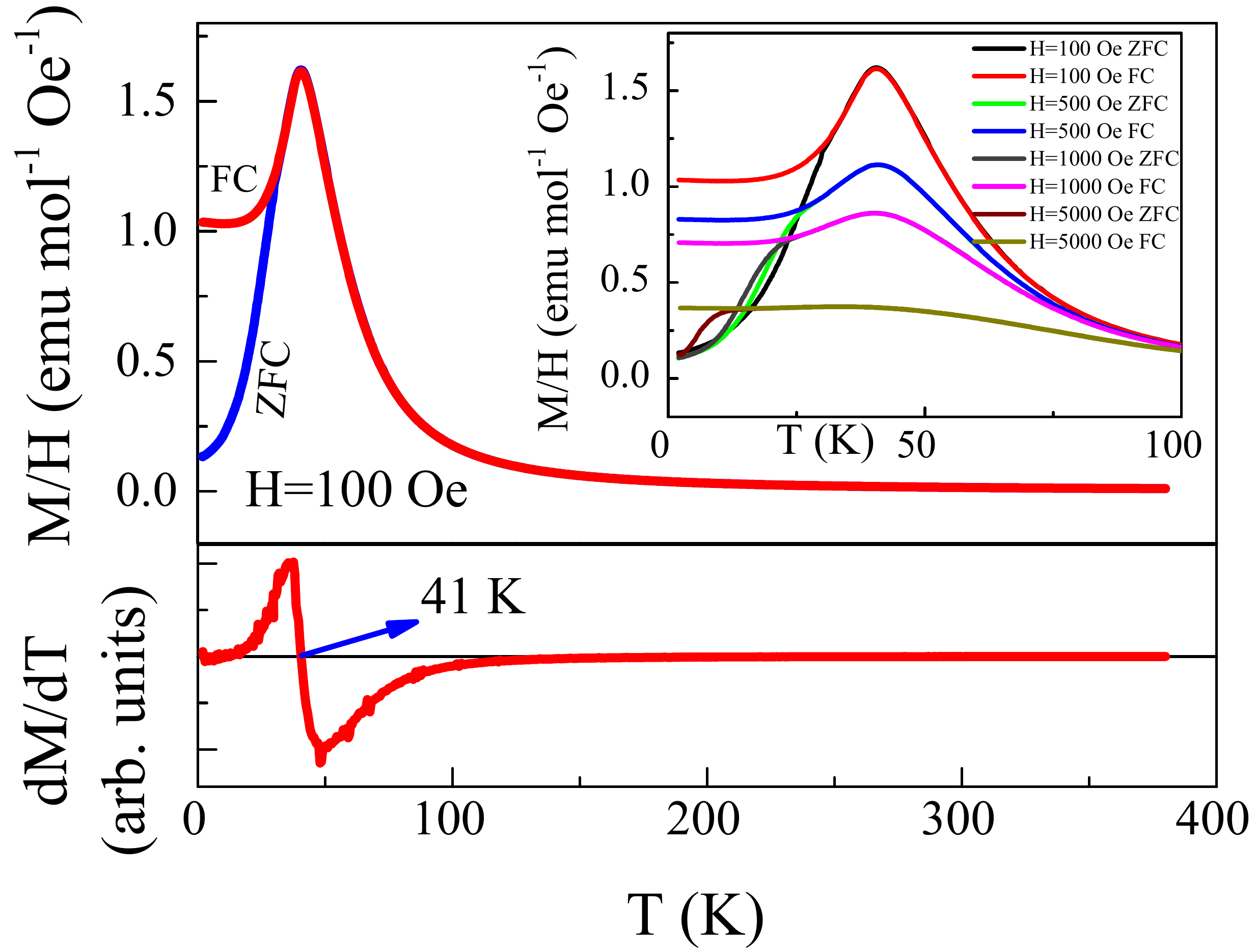}}
{\caption{(Upper panel) Temperature dependence of  magnetic susceptibility of FeRuMnGa measured in a 100 Oe applied magnetic field under zero-field-cooled (ZFC) and field-cooled (FC) protocols. (Lower panel) d${M}$/dT \textit{versus} T plot presented for FC mode. T$_{\rm N}$, corresponds to d$M$/dT = 0.}\label{Fig_2}}
\end{figure}

Fig.\ref{Fig_2} represents the temperature variation of the magnetic susceptibility of FeRuMnGa measured in the presence of 100 Oe magnetic field. The $\chi$(T) data measured in both ZFC and FC protocols start increasing  below 100 K followed by a clear broad peak around T$_{\rm P}$ $\sim$ 41 K. Such a peak is a typical characteristics of antiferromagnetic transition. The temperature derivative of the susceptibility shows crossover from positive to negative near T$_{\rm P}$ as well ( bottom panel, Fig.\ref{Fig_2}). Additionally, the $\chi$(T) recorded in FC protocol clearly shows nearly temperature invariant behavior and a bifurcation from ZFC data below an irreversibility temperature (T$_{irr}$), which was found to decrease with application of magnetic field (Fig.\ref{Fig_2}: inset of top panel) – a feature reminiscent with spin-glass (SG) like behaviour~\cite{pakhira2016large}. \\

\begin{figure}[h]
\centerline{\includegraphics[width=.48\textwidth]{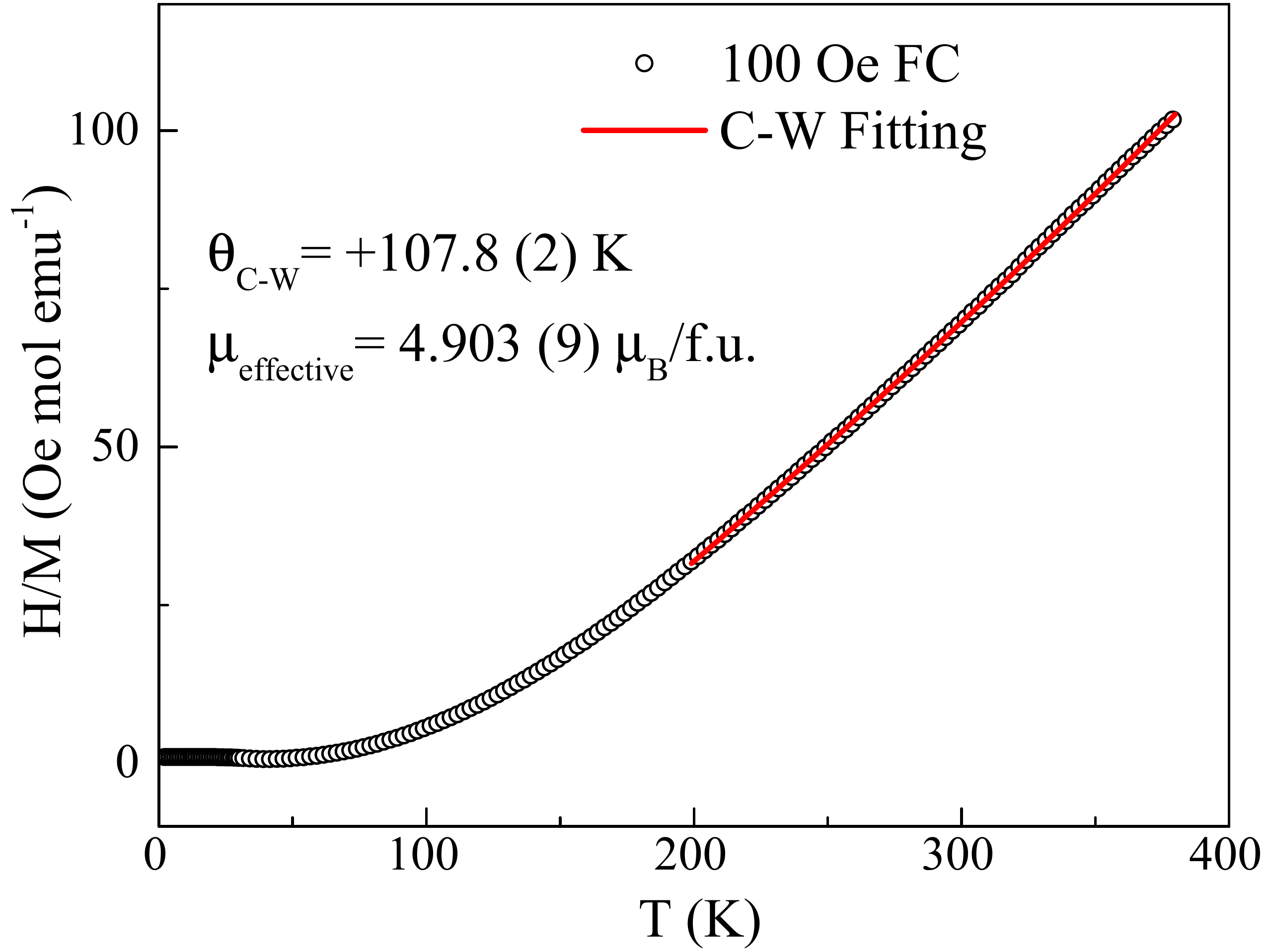}}
{\caption{Inverse magnetic susceptibility \textit{versus} temperature data recorded at 100 Oe in FC mode.}\label{Fig_3}}
\end{figure}

Curie-Weiss (C-W)~\cite{kundu2021complex} fit of the inverse susceptibility in the temperature region 200$-$380 K gives Curie-Weiss temperature (${\theta_{\rm CW}}$) = 107.8 K, which is nearly the same temperature below which both ZFC and FC susceptibility data started increasing in Fig~\ref{Fig_2}. The effective paramagnetic moment calculated from C-W fitting is $\sim$4.9 $\mu_{\rm B}$/f.u. (Fig.\ref{Fig_3}). The positive sign of ${\theta_{\rm CW}}$ indicates the nature of ground state magnetism of the compound to be of ferromagnetic nature.  However, this results is in contradiction with the observed antiferromagnetic-like transition at $\sim$41 K in the magnetic susceptibility data.

\begin{figure}[h]
\centerline{\includegraphics[width=.48\textwidth]{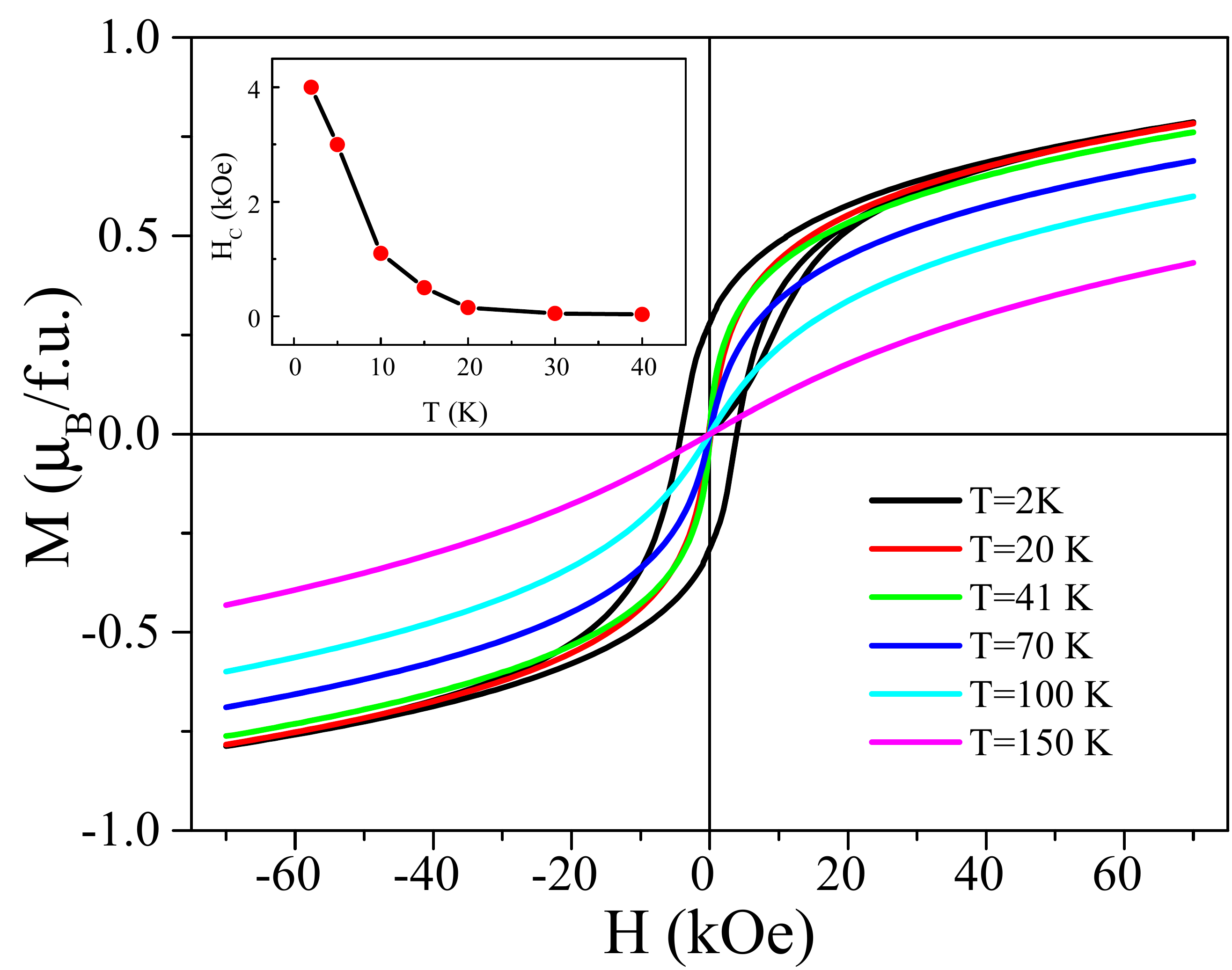}}
{\caption{Isothermal magnetization taken at different temperature in the range 2--150 K (for clarity some measured curves are not shown). Inset shows temperature variation of the coercivity (H$_C$).}\label{Fig_4}}
\end{figure}

The M(H) curve of the sample at 2 K exhibits a moderately large value of coercive field (H$_C$ $\sim$ 4 kOe) (Fig.~\ref{Fig_4}), which gradually diminishes with increasing temperature (Fig.~\ref{Fig_4}): inset). The manifestation of hysteresis in M(H) typically indicates the presence of ferromagnetic interaction in the sample. However, isothermal magnetization  does not saturate even at 2 K and reaches only a meager value of 0.80 $\mu_B$/f.u. at an applied field of 70 kOe, deviating largely from the ferromagnetic value of $\sim$ 2 $\mu_B$/f.u. expected according to the Slater-Pauling (S-P) rule~\cite{galanakis2002slater}, which also further rules out the presence of collinear ferromagnetic ground-state of the sample. Thus, from the M(H) and $\chi$ (T) behaviors, it can be concluded that the magnetic state of the sample at low temperature is neither true antiferromagnetic nor ferromagnetic. Furthermore,  the M (H) curve at high fields can roughly be considered to be consisting of two components: a linear component superimposed on a ferromagnetic-like saturation behavior, which may indicate that both ferromagnetic (FM) and antiferromagnetic (AFM) interaction co-exists in the magnetic ground state of the sample despite it manifests a antiferromagnetic-like transition in both ZFC and FC magnetization curves. The co-existence and competition  between competing FM and AFM states often leads to magnetic frustration promoting stabilization of spin glass-like state~\cite{mydosh1993spin}. It is worth mentioning that M(H) curve does not show non linear behavior even at much higher temperature than T$_P$ implying that a short-ranged magnetic correlation may exists even at high temperature.

\subsection{\label{sec:Heat Capacity}Heat Capacity}

\begin{figure}[h]
\centerline{\includegraphics[width=.48\textwidth]{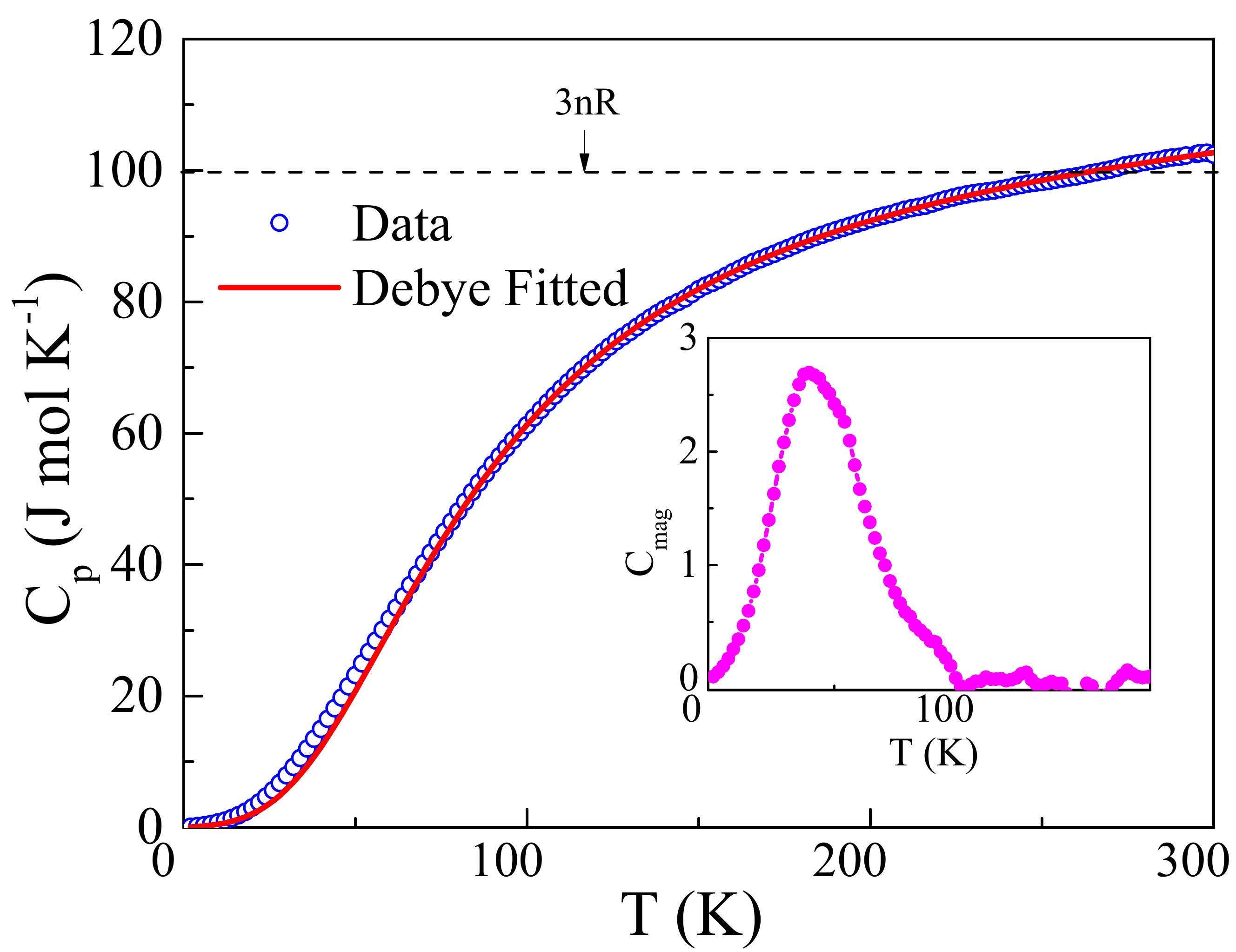}}
{\caption{Heat capacity (C$_P$) as a function of temperature. Inset shows magnetic contribution of the heat capacity (C$_P$) data. The hump in the experimental data near 290 K is due to melting of Apiezon N grease used in the measurement~\cite{design2002non}.}\label{Fig_5}}
\end{figure}

Heat capacity measurement is often used to confirm long-ranged magnetic transition in a compound, although many itinerant electron system are also known to suppress such signature. Fig.\ref{Fig_5} represents temperature variation of the heat capacity (C$_P$) of FeRuMnGa measured in absence of magnetic field. The room temperature value of the C$_P$ reaches to the classical limit predicted by Dulong-Petit,  which is \textit{3nR} where \textit{n} is the total number of atoms in the formula unit and is 4 for FeRuMnGa. The heat capacity data does not exhibit neither $\lambda$- nor $\delta$-like peak in the entire temperature range as expected in case of magnetic transition. We have attempted to find the lattice contribution of the heat capacity by fitting the heat capacity in the paramagnetic region (100$-$300 K) utilizing the standard Debye model~\cite{gopal2012specific} and extrapolating the fitted model down to 2 K. Magnetic contribution of the heat capacity (C$_{mag}$) can then be estimated by subtracting this phonon contribution from the measured C$_P$~\cite{singh2008observation,chatterjee2020glassy}. The resultant magnetic contribution, thus estimated, exhibits, a broad peak in the region 2-100 K with a maximum around 40 K (inset: Fig.~\ref{Fig_5}) , which is close to the temperature where $\chi$ (T) shows a peak (Fig.~\ref{Fig_2}). The manifestation of such broad peak in C$_{mag}$ has been ascribed as the spin-glass like transition in many other transition metal based itinerant magnetic systems, Mn$_3$In being prime example~\cite{chatterjee2020glassy}.

\subsection{\label{sec:Neutron}Neutron diffraction}

\begin{figure}[h]
\centerline{\includegraphics[width=.48\textwidth]{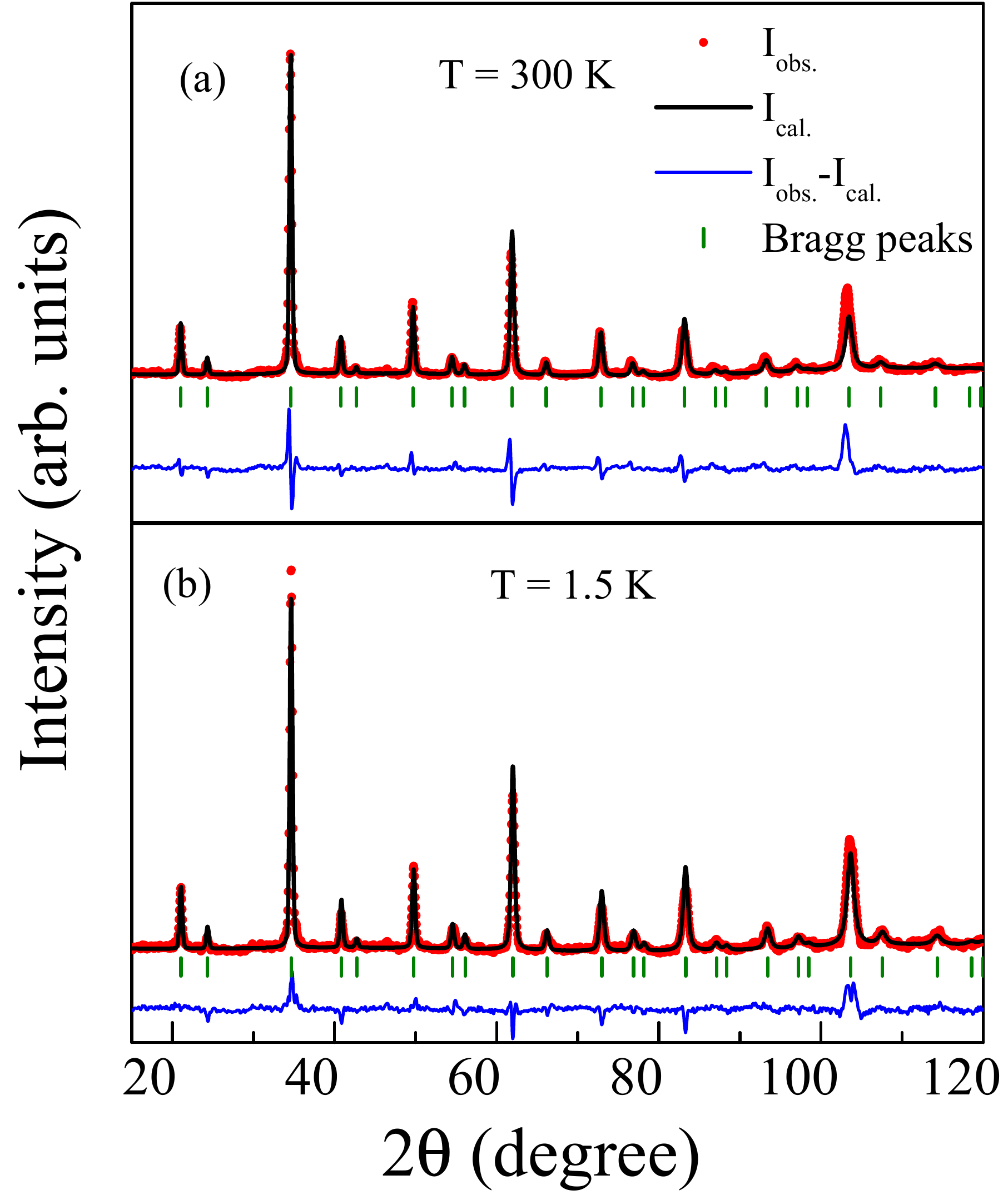}}
{\caption{Rietveld refinement of the neutron diffraction pattern of FeRuMnGa taken at (a) 300 K and (b) 1.5 K.}\label{Fig_6}}
\end{figure}

To get more insight into the magnetic ground state, we performed neutron diffraction (ND) study at 300 K (paramagnetic region) and 1.5 K (T $<$ T$_P$). B2-type of structural disorder model obtained from the Rietveld refinement of XRD data fails to explain the ND diffraction data taken at 300 K (Fig.~\ref{Fig_6} (a)). Interestingly, (111) peak is very prominent but the (200) peak is slightly diffused in nature indicating towards presence of a another kind of disorder in the studied compound rather than B2- and A2-type as discussed in Sec.~\ref{sec:XRD}. The structural disorder presented in Table\ref{Occupancy_NPD_Table} was assumed to yield the best fit. Due to the random variations of the scattering factors of the nearby elements from periodic table, neutron diffraction often founds very useful in determining the correct structure~\cite{kroder2019spin,mukadam2016quantification,ravel2002exafs,samanta2018reentrant} which is also the case here.

\begin{table}[]
\caption{Site occupancy of FeRuMnGa obtained from Neutron diffraction.}
\begin{tabular}{lll}
Site                                 & Element & Occupancy  (\%) \\\hline\hline
\multirow{3}{*}{4a (0,0,0)}          & Ga      & 47.6            \\
                                     & Mn      & 52.4            \\\hline
\multirow{3}{*}{4b (0.5,0.5,0.5)}    & Mn      & 14.1           \\
                                     & Ga      & 56.4            \\
                                     & Fe      & 29.5            \\\hline
\multirow{3}{*}{4c (0.25,0.25,0.25)} & Fe      & 34.7            \\
                                     & Ru      & 34.6            \\
                                     & Mn      & 30.7            \\\hline
\multirow{3}{*}{4d (0.75,0.75,0.75)} & Ru      & 67.9            \\
                                     & Fe      & 32.1            \\\hline
\end{tabular}
\label{Occupancy_NPD_Table}
\end{table}

Generally AFM compounds show additional peaks in the neutron diffraction data below their Neel temperature (T$_N$) while increase in the intensity for certain Bragg peaks are observed preferably at low angles for FM compounds. On the other hand, spin-glass systems often neither show additional magnetic peaks nor any increase
in the intensity of the Bragg peaks in the neutron diffraction pattern due to the absence of long-range order. Thus, the ND pattern taken at 1.5 K (Fig.~\ref{Fig_6} (b)) ($<$ T$_P$), which neither shows any additional peaks nor any increase in the intensity of the Bragg peaks, rules out the possibility of long-range magnetic ordering and suggests presence short-range magnetic ordering. We have analyzed the Rietveld refinement of the ND data taken at 1.5 K (Fig.~\ref{Fig_6} (b)) assuming the same structural model presented in Table~\ref{Occupancy_NPD_Table}.

Mn and Fe are the two magnetic ions present in the studied compound. Three kinds of magnetic interactions are possible \textit{viz.} Fe-Fe, Fe-Mn and Mn-Mn. In Heusler alloy containing Mn atoms, Mn-Mn interaction plays a major role in determining the nature of magnetism. Local moments of the Mn-atoms interact with nearest neighbours \textit{via} conduction electron through the oscillatory Ruderman Kittel-Kasuya-Yoshida (RKKY) exchange. Depending upon the distance between two Mn-atoms, the interaction becomes either positive (ferromagnetic) or negative (anti-ferromagnetic). Neutron diffraction suggests that in the studied compound Mn atoms are distributed in 3-sites (4\textit{a}, 4\textit{b} and 4\textit{c}). Due to this random distribution of Mn-atoms, FeRuMnGa lost its long range ordering and a magnetic frustration is expected due to the competing exchange interaction present in the system.

\subsection{\label{sec:ac_Susceptibility}ac susceptibility}

\begin{figure}[h]
\centerline{\includegraphics[width=.48\textwidth]{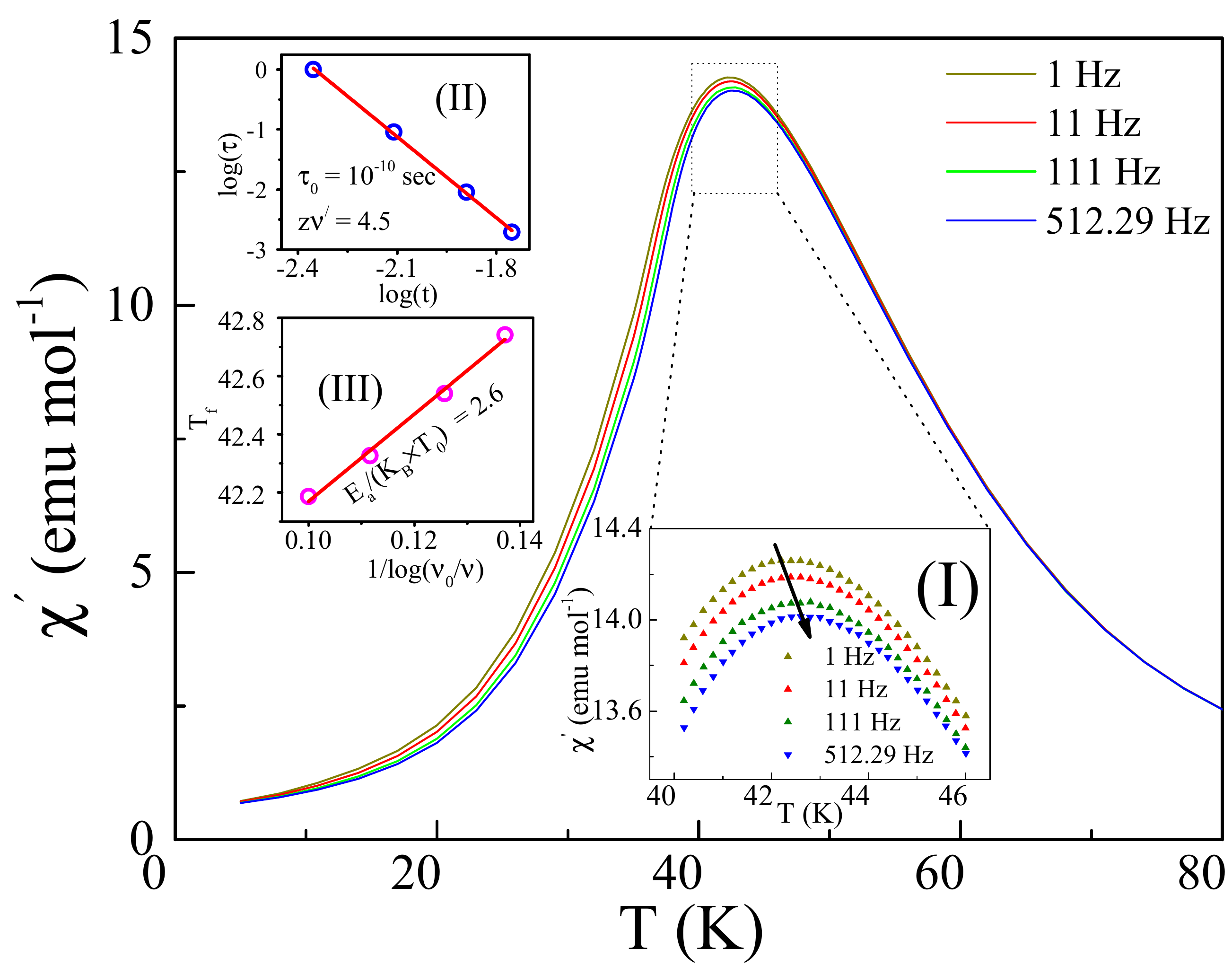}}
{\caption{Temperature dependence of the real part of the ac susceptibility of FeRuMnGa taken at different frequenies. The zoomed view of the frequency dependence in shown in inset (I). The frequency dependence of freezing temperature are shown in inset (II), where ln(t) are plotted as a function of ln(t), with t = (T$_f$ - T$_{SG}$/T$_{SG}$). The solid lines represent the fit to the power-law divergence. The frequency dependence of freezing temperature plotted as T$_f$ \textit{vs.} ln(f$_0$/f) is shown in inset (III). The solid line represents the fit to Vogel-Fulcher law.}\label{Fig_7}}
\end{figure}

The dc-magnetization, heat capacity and neutron diffraction studies suggests the lack of long ranged-magnetic ordering in ground state of the sample rather the stabilization of a spin-glass (S-G)  like state is more probable at low temperature. To confirm the occurrence of S-G like transition and associated magnetic dynamics, we have carried out detailed ac-susceptibility study on the sample. The ac-susceptibility data clearly shows a frequency-dependent peak around $\sim$42.5 K and above peak temperature all ac-susceptibility curves overlap each other. It is worth mentioning that dc-susceptibility data also shows a peak around that temperature (Fig.~\ref{Fig_2}). The shifting of peak towards high temperature with increasing frequency is a typical feature of spin-glass like transition and in that case the peak temperature corresponds to spin-freezing temperature (T$_f$) (Fig.~\ref{Fig_7}). The relative shift in freezing temperature per decade of frequency in a typical glassy system is commonly expressed as

\begin{eqnarray}
\delta T_{f} =\frac{\Delta T_{f}}{T_{f}\Delta (\log_{10} f)}
\end{eqnarray}
where \textit{f} is the frequency~\cite{mydosh1993spin}. The value $\delta T_{f}$ for canonical spin glasses have been reported to be $\sim$0.001, it is of the order of 0.01 for several spin cluster glass compounds~\cite{mydosh1993spin}, while the value is $\sim$0.1 for numerous known superparamagnetic systems. In the studied compound, $\delta T_{f}$ is estimated to be 0.004, which lies in between of canonical spin glass and cluser glass regimes. Similar information can also be extracted from the conventional power-law divergence of a critical slowing down equation where the frequency dependent shift of peak in ac susceptibility can be expressed as~\cite{mydosh1993spin,hohenberg1977theory}

\begin{eqnarray}
\tau = \tau_0\left(\frac{T_f-T_{SG}}{T_{SG}}\right)^{-z\nu^{\prime}}\label{eqn:dynamical scaling}
\end{eqnarray}
\noindent
where $\tau$ is the relaxation time associated with the measured frequency ($\tau$=1/\textit{f}), $\tau_0$ is the single-flip relaxation time, T$_{SG}$ is the spin-glass temperature for \textit{f} = 0, and z$\nu^{\prime}$ is the dynamical critical exponent. The value of z$\nu^{\prime}$ typically lies between 4-12 for spin glass state. The value of $\tau_0$ for canonical spin glasses is in the region of  $10^{-13}$ -- $10^{-12}$, but the value of $\tau_0$ for a spin cluster glass system is typically in the range of $10^{-11}$ -- $10^{-4}$~\cite{lago2012three,mori2003dynamical,chakraborty2022ground}. Superparamagnetic state development is associated with larger values of $\tau_0$ . For the present FeRuMnGa, the value of z$\nu^{\prime}$ is found to be 4.5 which is in the range spin glass state formation and $\tau_0$ = $10^{-10}$ secs which also lies in the border range between canonical and cluster glass states. Another dynamical scaling law, known as the Vogel-Fulcher relation, can be used to simulate spin dynamics in glassy systems around the freezing temperature. According to Volgel-Fulcher (V-F) relation, frequency dependence can be expressed as~\cite{mydosh1993spin,souletie1985critical}
\begin{eqnarray}
f = f_{0} \exp \left[-\frac{E_a}{k_B(T_f-T_0)}\right]\label{eqn:Vogel-Fulcher}
\end{eqnarray}
\noindent
where $f$ = 0 is known as the characteristic attempt frequency, E$_a$ is the activation energy and T$_0$ is the Vogel-Fulcher temperature. From the $T_f$ \textit{versus} 1/$\log\frac{f_0}{f}$ plot for FeRuMnGa, the fitted values are found to be $E_a$/K$_B$ = 40.2 and T$_0$ = 14.9. For canonical spin glass state the value of $\frac {E_a} {K_{B}T_{0}}$ is reported to be close to 1 whereas for cluster glass type of system this value is relatively larger. In the studied compound, the value of $\frac {E_a} {K_{B}T_{0}}$ is found to be 2.6 which belongs to the cluster glass regime.  Thus from the ac-susceptibility study, it can be inferred that the magnetic state below 40  K for the sample is borderline between canonical spin glass and cluster glass, but is closer to the latter.

\subsection{Magnetic relaxation}

\begin{figure}[h]
\centerline{\includegraphics[width=.48\textwidth]{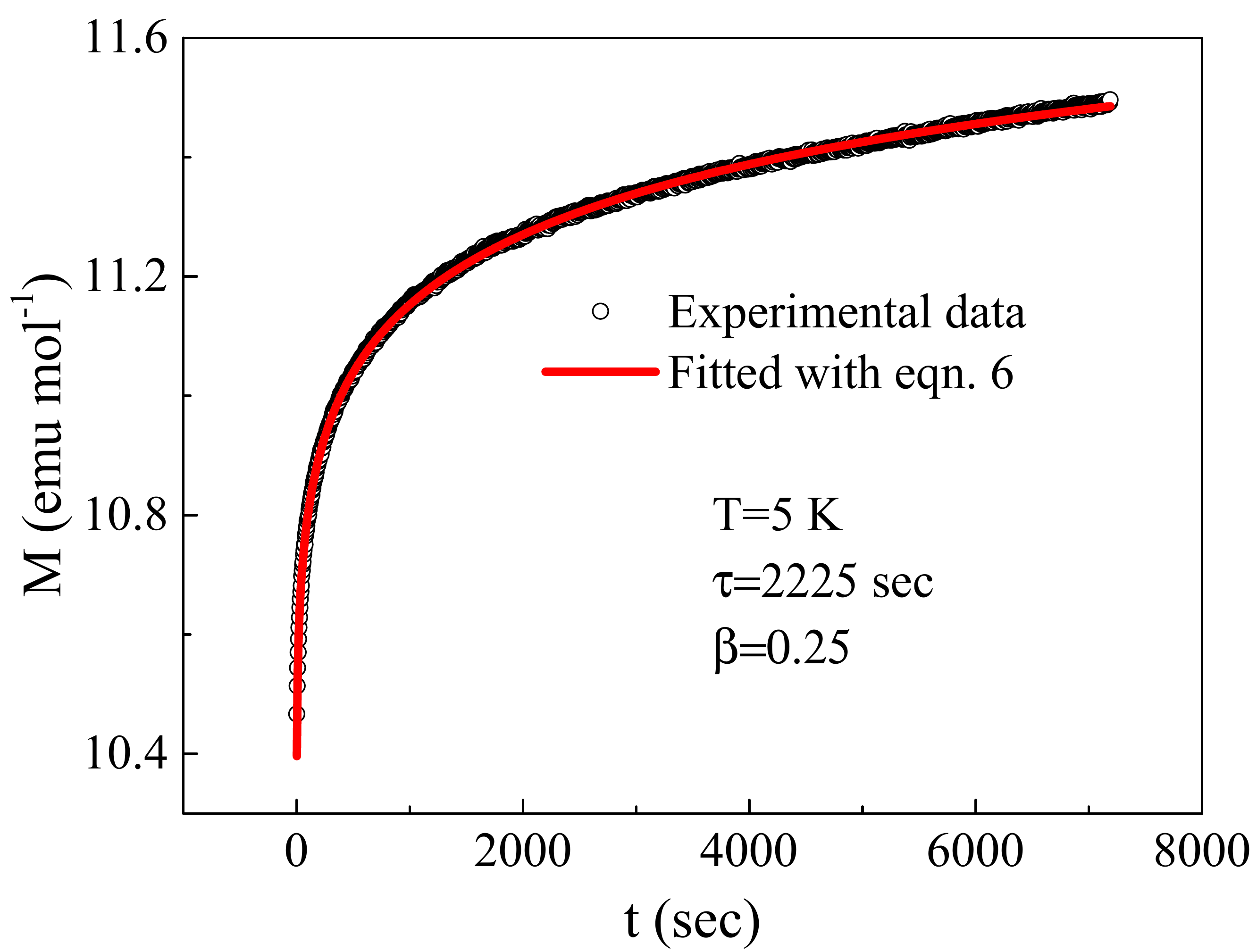}}
{\caption{Time dependent magnetization data of FeRuMnGa taken at T=5 K under zero field cooled condition.}\label{Fig_8}}
\end{figure}

 To get insights into the glassy behavior, we carried out magnetic relaxation study. Magnetic relaxation behavior was measured in zero-field-cooled (ZFC) mode, where the sample was cooled from paramagnetic region to the measurement temperature  T = 5~K ($ < T_f$), in the absence of any magnetic field. After the temperature stabilization for a wait time (t$_w$), a small amount of  magnetic field (H) of 100 Oe was applied and the time dependency of the magnetization $M(t)$ was recorded as shown in Fig.~\ref{Fig_8}. A clear magnetic relaxation behavior is observed where the $M(t)$ asymptotically approaches saturation over a long time-scale following the empirical stretched-exponential function of the form~\cite{pakhira2018chemical,mondal2020non},

\begin{equation}
M(t) = M_{0} + M_{g}\exp \left[-\left(\frac{t}{\tau}\right)^{\beta}\right]\label{eqn:relax}
\end{equation}
\noindent
where M$_0$ is intrinsic magnetization, M$_g$ is the glassy component of magnetization,  $\tau$  is the relaxation time and  $\beta$ is known as the stretching exponent. The value of  $\beta$ varies within 0 to 1 for different spin glass systems depending on the nature of energy barriers associated with the spin-glass state~\cite{mydosh1993spin,chu1994dynamic}. $\beta$ = 0 rules out any possibility of relaxation behavior, whereas, $\beta$ = 1 signifies the presence of single time-constant relaxation process. For the studied compound the value of  $\beta$ and  $\tau$  are found to be = 0.25 and 2225 sec, respectively, which are in the similar range to that of different earlier reported spin glass systems~\cite{chatterjee2020glassy,pakhira2016large,kroder2019spin}.

\subsection{Magnetic memory effects}
\begin{figure}[h]
\centerline{\includegraphics[width=.48\textwidth]{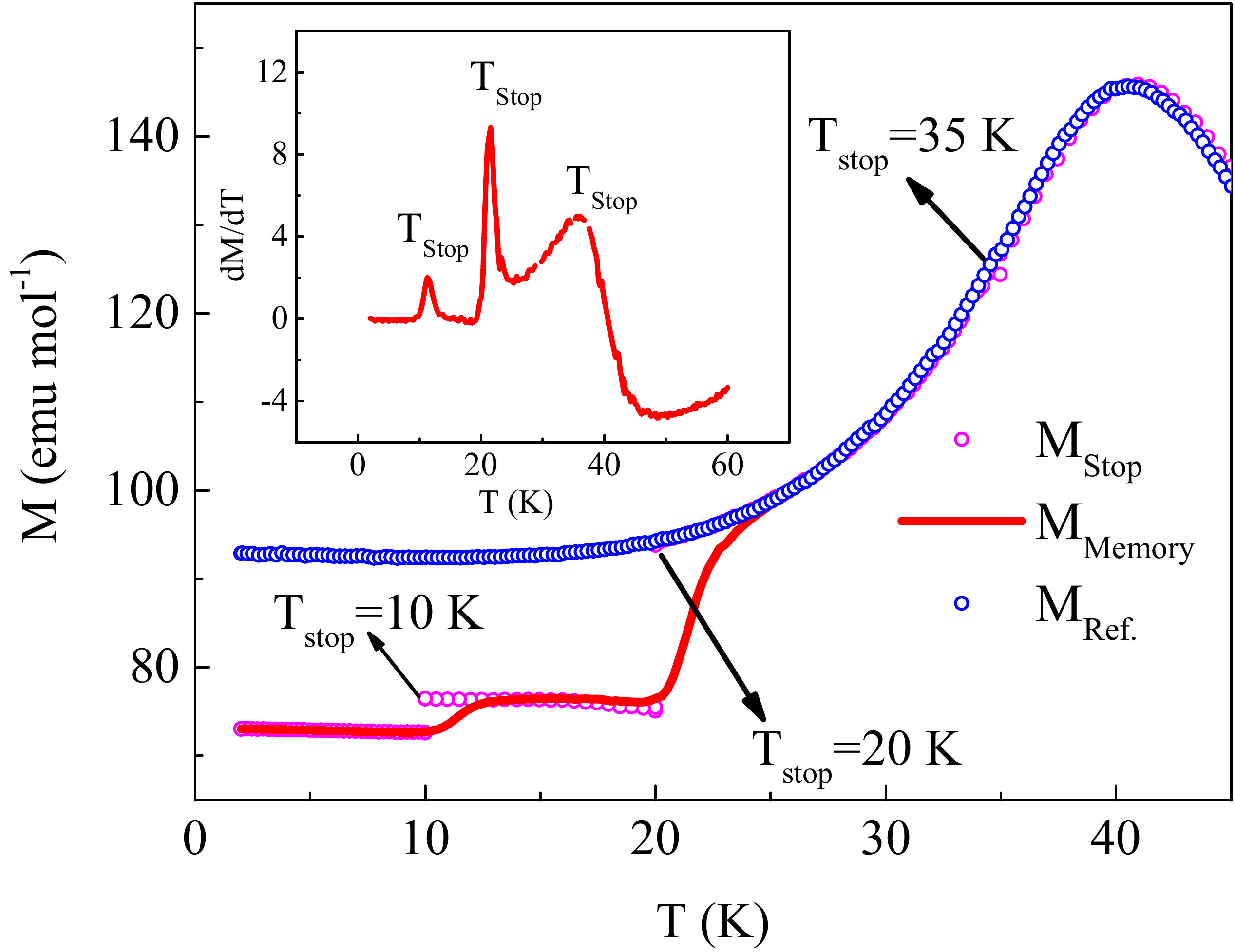}}
{\caption{Memory effect in the FC condition.}\label{Fig_9}}
\end{figure}
Beside magnetic relaxation, magnetic memory effect is another salient feature of the spin glass state~\cite{pakhira2016large,ghara2014reentrant,bhattacharyya2011spin}. Field-cooled (FC) magnetic memory measurement was performed for the studied compound following the protocol described by Sun\textit{ et. al.}~\cite{sun2003memory}. The sample was initially cooled from the paramagnetic region under 100 Oe applied magnetic field and upon reaching the stopping temperatures (T$_{Stop}$) of 35 K, 20 K and 10 K ($< T_f$), the magnetic field switched off at each temperature for a duration of t$_w$ = 1.5~h. After the lapse of  t$_w$, the magnetic  field was turned on with resumed cooling. Temperature dependence of magnetization recorded in this process is depicted as $M_{FC}^{Stop}$, as shown in Fig.~\ref{Fig_9}. After reaching the lowest measurement temperature of 2 K, the sample was measured on heating to the paramagnetic region without any stop. The M(T) behavior recorded is this process is M$_{FC}^{Mem}$. A conventional field-cooled magnetization response is also recorded and referred to as the reference magnetization M$_{Ref}$, as shown in Fig.~\ref{Fig_9}. Magnetic memory in this FC process is clearly evidenced in the compound as shown in Fig.~\ref{Fig_9}, where M$_{FC}^{Mem}$ tries to follow the $M_{FC}^{Stop}$ behavior yielding an anomaly bending at each T$_{Stop}$. This observation signifies that the system remembers it's previous state history. Presence of such FC memory effect is typical in different spin glass systems associated with the non-equilibrium time-dependent magnetization dynamics~\cite{kroder2019spin,chatterjee2020glassy,pakhira2016large,mondal2019physical}.

\begin{figure}[h]
\centerline{\includegraphics[width=.48\textwidth]{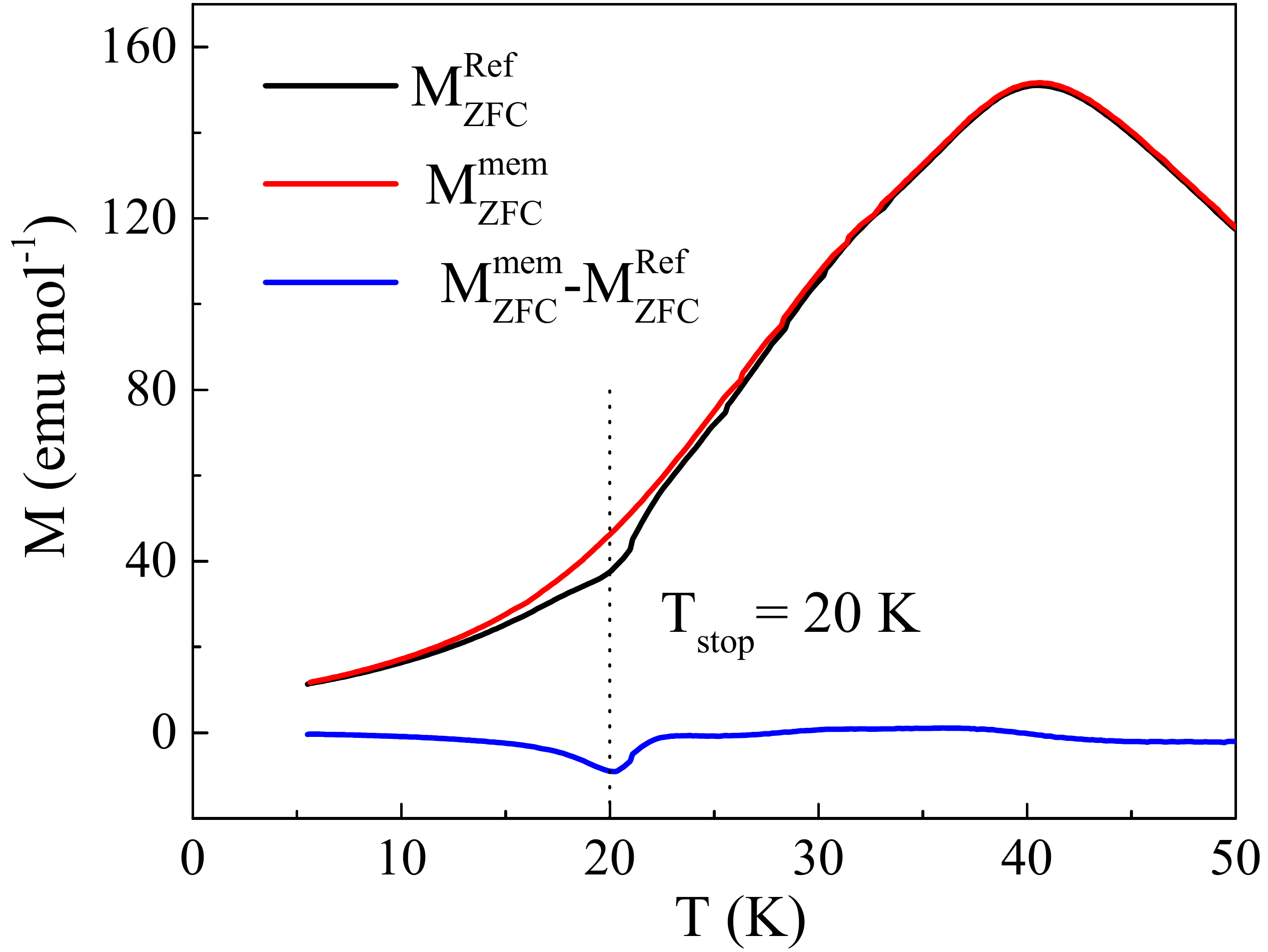}}
{\caption{Memory effect in the ZFC condition.}\label{Fig_10}}
\end{figure}

The memory effect under the zero-field-cooled (ZFC) protocol was also studied in the present compound. In the ZFC protocol, the sample was first cooled down at zero-field from the paramagnetic region to the stopping temperature T$_{stop}$ = 20 K, where  the temperature was hold for a wait time t$_w$ = 1.5 h. Then the sample was again cooled to the lowest measurement temperature of 2 K. The magnetization M(T) was then recorded during heating from 2 K to the paramagnetic region under the application of a 100 Oe magnetic field. The M(T) curve obtained in this process is labelled as M$_{ZFC}^{Mem}$. The reference ZFC magnetization for the 100 Oe field is also measured without any temperature halt. This is designated as M$_{ZFC}^{ref}$. The ZFC memory effect of the studied compound is shown in Fig.~\ref{Fig_10}, where the difference in magnetization, $\delta$M = M$_{ZFC}^{Mem}$ - M$_{ZFC}^{ref}$, exhibits a clear memory dip around the stopping temperature, indicating the presence of ZFC memory effect.

It may be pointed out here that the memory effect is also observed in superparamagnetic systems in the FC process. Only the ZFC memory effect can differentiate the spin glass class from a superparamagnetic system as superparamagnetic compounds do not show a memory effect in the ZFC protocol~\cite{sasaki2005aging}. Thus, the observed memory effect in ZFC mode confirms the formation of a spin glass state in the studied compound.

\begin{figure}[h]
\centerline{\includegraphics[width=.48\textwidth]{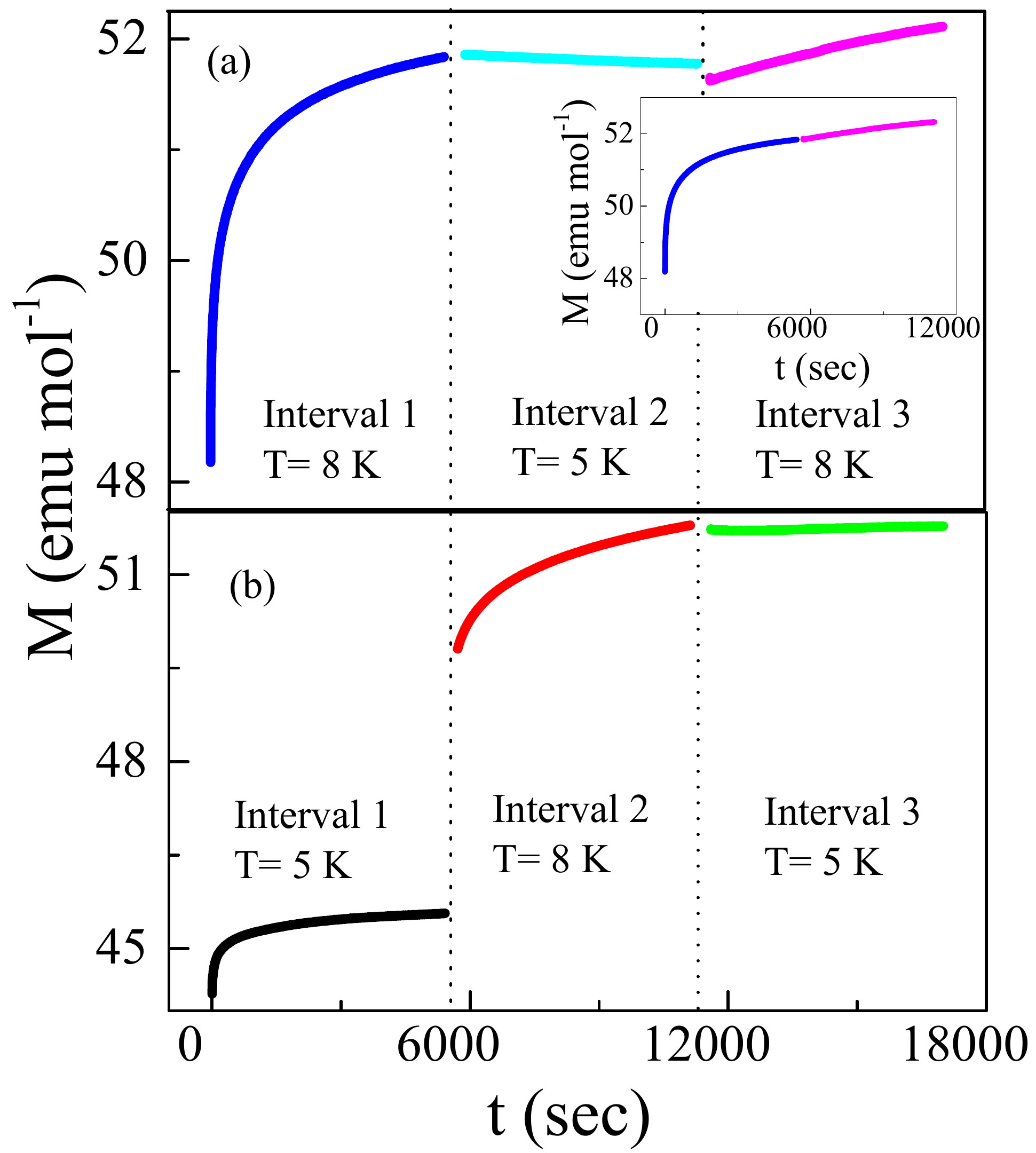}}
{\caption{Memory effect taken (a) in the intermediate cooling cycle (b) in the intermediate heating cycle. Inset shows merging of the interval 1 and interval 3 data for the intermediate cooling cycle.}\label{Fig_11}}
\end{figure}

The droplet~\cite{mcmillan1984scaling,fisher1988nonequilibrium} and the hierarchical models~\cite{vincent1997slow,ulrich2003slow} are two widely used theoretical models to describe the memory behavior in different spin-glass systems. The droplet model deals with uniform spin configuration, whereas the hierarchical model predicts a multivalley free-energy landscape with multiple potential spin configurations at a certain temperature. As a result of that, during a temperature cycling, the hierarchical model only predicts the observation of the memory effect for intermediate cooling, while the droplet model predicts memory effect for both heating and cooling protocols. In order to verify which model is applicable in the present case, we have studied the memory effect in both the above mentioned protocols by  Sun \textit{et al}~\cite{sun2003memory}. At first, the sample was zero-field-cooled from the paramagnetic state to T = 8 K, than a magnetic field of 100 Oe was applied and M(t) was recorded for t = 6,000 s (interval 1). Then the temperature was suddenly lowered to 5 K and M(t) was measured for another t = 6,000 s at that fixed temperature (interval 2). Finally, the temperature was again increased to 8 K (interval 3) followed by a M(t) measurement for t = 6,000 s. The measured M(t) behavior in this whole process are shown in Fig.{~\ref{Fig_11}(a)}. The magnetization data from intervals 1 and 3 may be combined to show that both branches fit as if no intermediary cooling had occurred. After warming, the system \enquote{memorises} its previous condition before the interim cooling. An inverse-temperature cycling was also applied to study the temporary heating effect, as shown in Fig.{~\ref{Fig_11}(b)}. The only modification to the earlier process is the intermediate heating instead of intermediate cooling. In this case, the magnetization does not revert to the value it had prior to the intermediate heating.  Since the memory effect is only seen during intermediate cooling, the hierarchical model is applicable in the studied system, in agreement with many other reported spin-glass systems~\cite{kroder2019spin,pakhira2016large,chakrabarty2014cluster}.

\subsection{\label{sec:Resistivity}Resistivity}

\begin{figure}[h]
\centerline{\includegraphics[width=.48\textwidth]{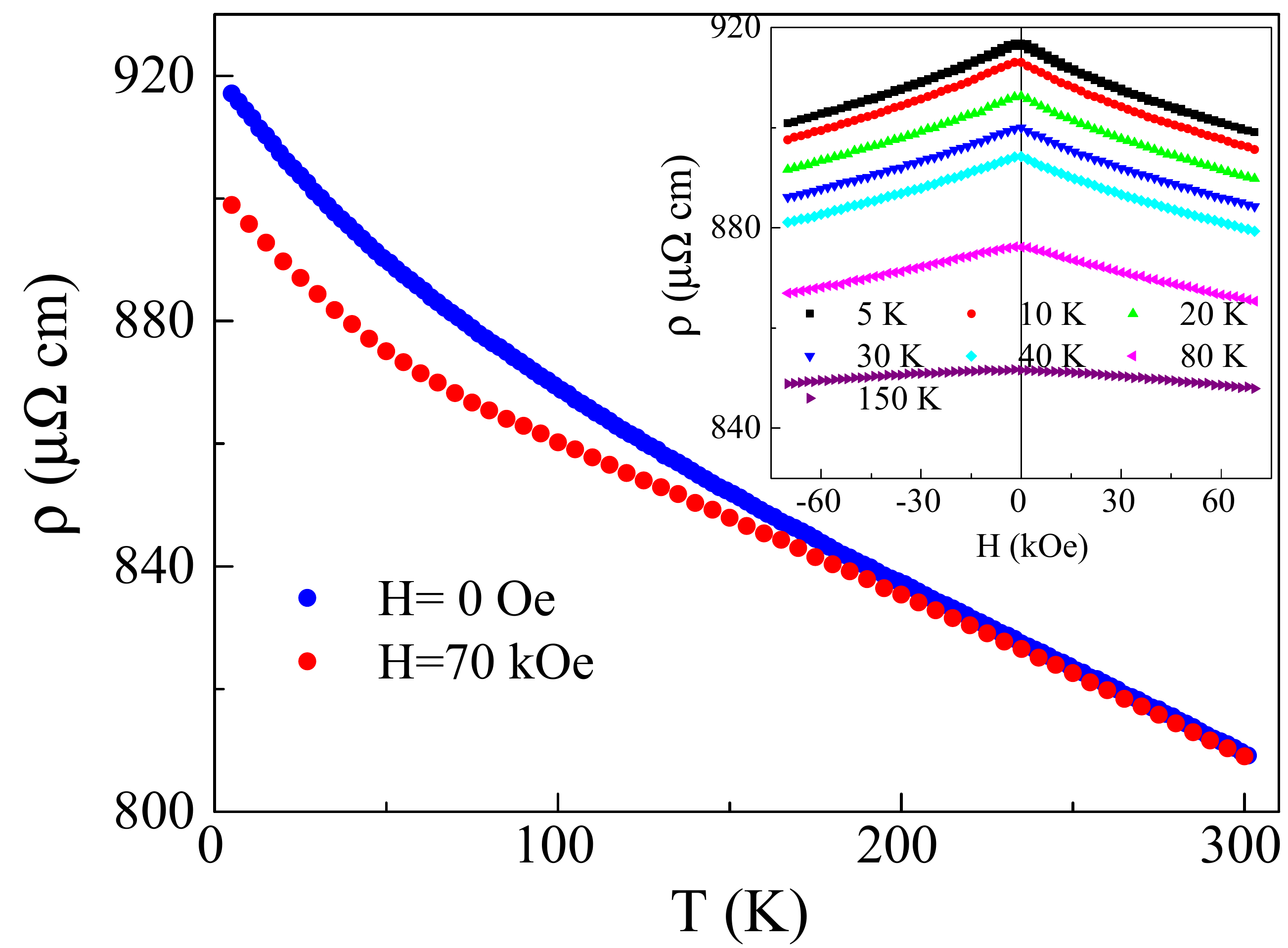}}
{\caption{Temperature dependence of the electrical resistivity measured in the absence of magnetic field in the temperature range 5$-$300 K. Inset shows magnetoresistance taken at different temperatures}\label{Fig_12}}
\end{figure}

To find the impact of the glassy magnetic state in the electrical transport properties, we have measured the longitudinal resistivity ($\rho_{xx}$) in zero field in both cooling and warming modes and found that no thermal hysteresis is present in the studied compound ruling out the presence of any structural changes. Temperature variation of the resistivity data taken at zero field warming mode is presented in Fig.~\ref{Fig_12}. The temperature variation of the resistivity data shows negative temperature coefficient behaviour throughout the whole measured temperature range. These type of negative temperature coefficient is typical for a disordered material and was earlier observed for other Heusler alloys~\cite{kaveh1982universal,lue2004thermal,mondal2018ferromagnetically,kroder2019spin}. The temperature variation of the resistivity data could not be fitted neither with the activated type of electrical transport behavior nor with the variable range hoping (VRH) conductivity models, which are usually used to explain the semiconducting nature of the resistivity observed for other Heusler alloys~\cite{mondal2018ferromagnetically,chatterjee2020glassy}. We have also measured temperature variation of the resistivity at 70 kOe (data presented in Fig.~\ref{Fig_12}). No sharp or abrupt change in resistivity was observed at the spin freezing temperature. Similar type of feature was also reported earlier for IrMnGa~\cite{kroder2019spin}. As short ranged magnetic correlations exists in much higher temperature even at $\sim$150 K as evidenced  through non-linear M(H) (Fig.~\ref{Fig_4}), the application of magnetic field suppresses the resistivity from  much higher temperature than T$_P$ by minimizing the spin disorder. This is consistent with the results discussed in isothermal magnetization measurement taken above freezing temperature. The minor change of the resistivity in presence of the field was also evident from the magnetoresistance (MR) measurements presented in the inset of Fig.~\ref{Fig_12} taken at different temperatures. The maximum MR measured at 5 K is found to be -1.88\%  under application of 70 kOe.

\subsection{\label{sec:Seebeck}Seebeck coefficient and Hall resistivity }
\begin{figure}[h]
\centerline{\includegraphics[width=.48\textwidth]{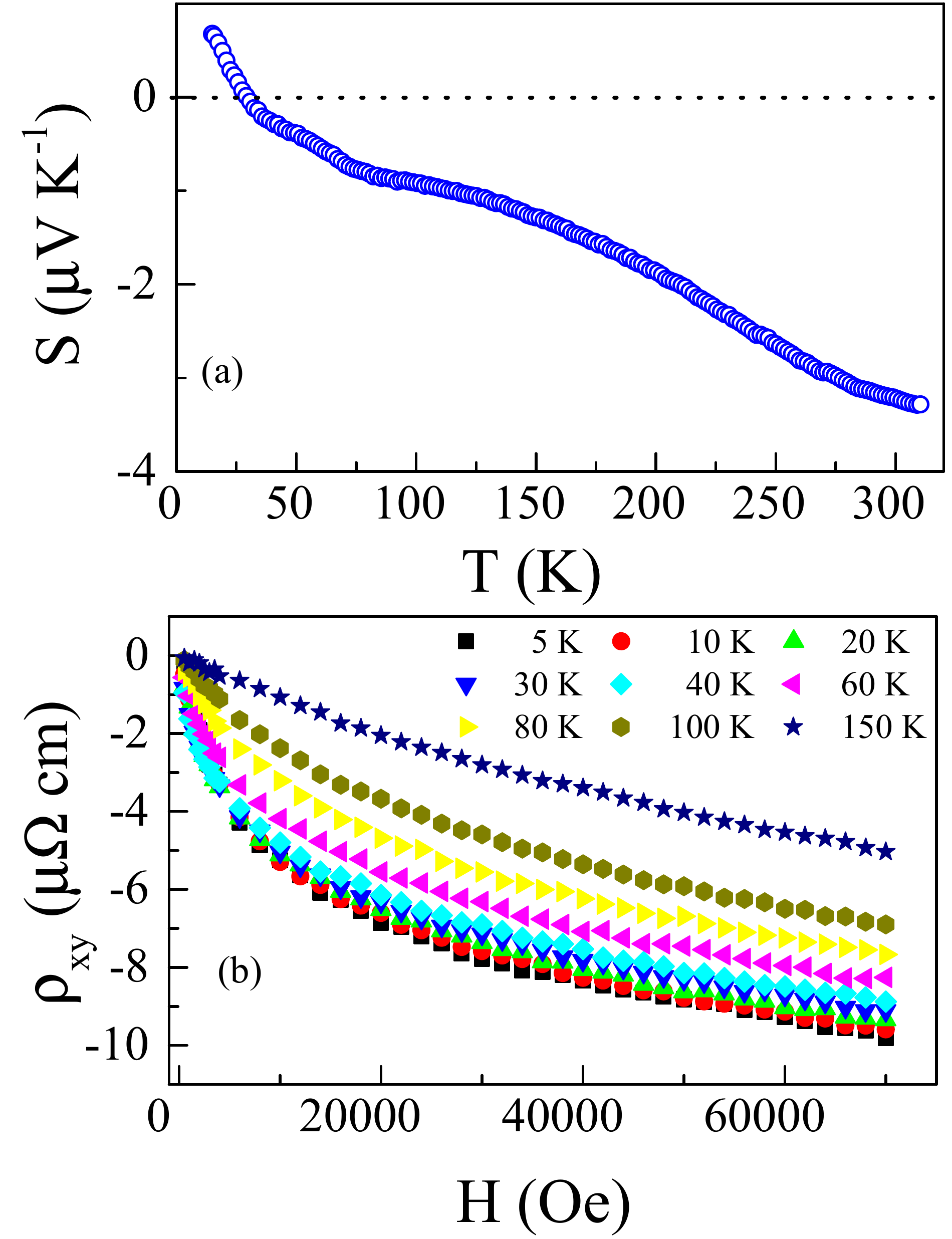}}
{\caption{ (a) Temperature dependence of the Seebeck coefficient measured in the absence of magnetic field in the temperature range of 15$-$310 K. (b) Hall resistivity ($\rho_{xy}$) \textit{versus} H measurements taken at different temperatures.}\label{Fig_13}}
\end{figure}

To get deeper understanding of electrical transport behavior, we also carried out thermopower and Hall resistivity studies. {Fig.~\ref{Fig_13}(a)} represents the temperature variation of the Seebeck coefficient measured within the range of 15-310 K. Seebeck coefficient is negative indicating electrons as the majority carriers in the studied sample.  The overall value of Seebeck coefficient is found to be quite small, S = 3.22 $\mu$V/K at 300 K. Additionally, there is a crossover from negative to positive values of S near T = 30 K. This crossover temperature is lower than the observed magnetic spin-freezing temperature (T$_f$ $\sim$ 41 K). Generally, in simplified Drude-Sommerfeld model Seebeck coefficient is defined as

\begin{equation}
S(T ) = \frac{8\pi^2{k_B}^2T}{3eh^2}m^*\big({\frac{\pi}{3n}})^{2/3}
\end{equation}
where \textit{e} is the electronic charge, \textit{n} is the density of the charge carriers, \textit{m$^*$} is the effective mass, \textit{k$_B$} is the Boltzmann constant  and  \textit{h} is Planck's constant~\cite{snyder2011complex}. Normally, the change in sign of the Seebeck coefficient is associated with the change in carrier type. The crossover from positive to negative near $\sim$ 30 K can be associated with the change of the majority carriers from electrons to holes. It is better to mention that simplified Drude-Sommerfeld model predicts linear variation of the Seebeck coefficient with the temperature. In the studied compound Seebeck coefficient does not show linear temperature dependence. To confirm the change of the carrier type observed in the S \textit{versus} T data, we have performed the Hall measurements at different temperatures. Temperature variation of the Hall resistivity ($\rho_{xy}$) taken at different temperatures is represented in the Fig.~\ref{Fig_13}(b). As can be clearly observed the $\rho_{xy}$ for all the measured temperatures lies in the negative region which indicates that electrons are the majority charge carriers for the studied compound consistent with the Seebeck results. The $\rho_{xy}$ \textit{versus} H data taken at different temperatures mimics the isothermal magnetization taken at different temperatures (Fig.~\ref{Fig_4}).  Hall resistivity can be described  as $\rho_{xy}$(T) = ${\rho^{OHE}_{xy}}$ + ${\rho^{AHE}_{xy}}$ = R$_0$H + R$_A$M, where ${\rho^{OHE}_{xy}}$ and ${\rho^{AHE}_{xy}}$ are the ordinary and anomalous hall contributions, respectively, and R$_0$, R$_A$, and M are the ordinary, anomalous Hall coefficient and magnetization, respectively~\cite{roy2020anomalous}. Ordinary Hall coefficient is linearly proportional to H and  anomalous Hall coefficient is proportional to magnetization of the sample. The Hall resistivity remains non-linear even upto 70 kOe. Anamalous Hall effect dominates over the ordinary Hall effect in the studied compound. In spin-glass state the generation of anomalous Hall effect is explained with the non-coplanar spin structure of the frustrated spin~\cite{pureur2004chiral}. Anamalous Hall effect for spin-glass state was earlier observed in half-Heusler IrMnGa~\cite{kroder2019spin}. We have not found any change of the carrier type from the Hall measurement which is earlier evident in the Seebeck results. This type of discrepancy between the Seebeck and Hall results was earlier observed for Mn$_3$In~\cite{chatterjee2020glassy}. The intricate details of electron transport properties of this highly disordered Heusler alloys  will be focus in our future study.

\section{Conclusion}
We successfully synthesized a equiatomic FeRuMnGa, a quaternary Heusler alloy with highly disordered structure in which two (Fe, Mn) of its magnetic constituent elements are distributed in three sites. The sample shows clear spin glass behavior at low temperature which is probed through dc magnetization, ac susceptibility, and magnetic memory experiments in combination with neutron diffraction study. Our detailed analysis of ac-susceptibility data reveals the magnetic state at low temperature in border-line of canonical spin glass and cluster glass. The effect of structural disorder is also reflected in the transport properties as the temperature dependence of resistivity exhibits nonmetallic character. Combined Seebeck and Hall resistivity data confirms that electrons are the majority charge carriers in the studied compound. However, the Seebeck coefficient suggests a change of the carrier type near 30 K, although such signature could not be verified through Hall resistivity data. The anomalous Hall contribution completely dominates the Hall resistivity.

\section{Acknowledgement}
We would like to dedicate the paper in the memory of  senior co-author: Late Prof. Vitalij. K. Pecharsky who recently passed away before submission of the paper.  S.G and S.C would like to sincerely acknowledge SINP, India and UGC, India, respectively, for their fellowship. Work at the Ames National Laboratory was supported by the Division of Materials Science and Engineering, Basic Energy Sciences, Office of Science of the US Department of Energy (US DOE). Ames National Laboratory is operated for the US DOE by Iowa State University of Science and Technology under Contract No. DE-AC02-07CH11358.
 %\newpage
\normalem
\bibliographystyle{apsrev4-2}
%\bibliography{reference}
%

\end{document}